\documentclass[a4paper,11pt]{article}

\usepackage[utf8]{inputenc}
\usepackage[T1,T2A]{fontenc}
\usepackage{amsmath,amsfonts,amssymb}

 \usepackage{graphicx}
\usepackage{wrapfig}
\usepackage{subfig}
\usepackage{hyperref}
\usepackage[dvipsnames]{xcolor}

\usepackage{esvect}

\begin{document}

\begin{titlepage}

\vspace{1.5cm}

\begin{center} {\LARGE \bf  Preheating with deep learning} \end{center}

\vspace{1cm}

\begin{center}
  {\bf Jong-Hyun Yoon, Simon Cl\'ery, Mathieu Gross and Yann Mambrini  }
\end{center}
  
\begin{center}
  \vspace*{0.15cm}
  \it{Universit\'e Paris-Saclay, CNRS/IN2P3, IJCLab, 91405 Orsay, France
 }
\end{center}
  
\vspace{2.5cm}

\begin{center} {\bf Abstract} \end{center}
 
 \noindent  We apply deep learning techniques to the late-time turbulent regime in a post-inflationary model where a real scalar inflaton field and the standard model Higgs doublet interact with renormalizable couplings between them. After inflation, the inflaton decays into the Higgs through a trilinear coupling and the Higgs field subsequently thermalizes with gauge bosons via its $SU(2)\times U(1)$ gauge interaction. Depending on the strength of the trilinear interaction and the Higgs self-coupling, the effective mass squared of Higgs can become negative, leading to the tachyonic production of Higgs particles. These produced Higgs particles would then share their energy with gauge bosons, potentially indicating thermalization. Since the model entails different non-perturbative effects, it is necessary to resort to numerical and semi-classical techniques. However, simulations require significant costs in terms of time and computational resources depending on the model used. Particularly, when $SU(2)$ gauge interactions are introduced, this becomes evident as the gauge field redistributes particle energies through rescattering processes, leading to an abundance of UV modes that disrupt simulation stability. This necessitates very small lattice spacings, resulting in exceedingly long simulation runtimes. Furthermore, the late-time behavior of preheating dynamics exhibits a universal form by wave kinetic theory. Therefore, we analyze patterns in the flow of particle numbers and predict future behavior using CNN-LSTM (Convolutional Neural Network combined with Long Short-Term Memory) time series analysis. In this way, we can reduce our dependence on simulations by orders of magnitude in terms of time and computational resources.

\end{titlepage}

\tableofcontents

\section{Introduction}

Cosmic inflation has been demonstrated to play a pivotal role in establishing the initial conditions of the early universe within modern cosmology \cite{Starobinsky:1980te, Guth:1980zm,Linde:1981mu,Mukhanov:1981xt}. These characteristics can be realized by a real scalar field, known as the inflaton. It allows for the exploration of various intriguing phenomenologies due to the interplay between the inflaton and other fields \cite{Lebedev:2021zdh, Lebedev:2021ixj, Lebedev:2022vwf}. However, in any post-inflationary model, we must eventually explain a thermal universe where the constituents of thermal plasma possess distinctive properties such as thermal mass, necessitating a deeper comprehension of thermal field theory. During the initial stages of thermalization, defining thermal quantities becomes challenging, particularly with a low number of plasma constituents \cite{Kurkela:2011ti}. In the initially under-occupied systems, low-momentum particles form a bath, thermalize, and then dominate the system by absorbing the energy from the hard excitations. This process is dominated by number-changing interactions, which can interfere and therefore it is important to take into account modification of the interaction rate so-called the Landau–Pomeranchuk–Migdal (LPM) effect \cite{Landau:1953um,Migdal:1956tc,Gyulassy:1993hr}. When the rate of plasma constituent production is sufficiently low, interaction dynamics can be treated perturbatively \cite{Harigaya:2013vwa, Mukaida:2024jiz}, but when the rate is large, the system falls into the non-perturbative regime, where fields redistribute their energies by turbulent rescattering. This late-time phenomenon is known as free turbulence, where the dynamics are described by kinetic theories \cite{Micha:2002ey, Micha:2004bv}. Given the inherent challenges of demonstrating free turbulence and tracking thermalization using lattice simulations, it is tempting to infer patterns in the occupation number of fields, represented as self-similarity implied by the kinetic theories and analyze them using deep learning methods for its proficiency in pattern recognition.

Machine learning and deep learning have revolutionized various fields, including physics, by providing powerful tools to analyze complex systems and phenomena \cite{LeCun:2015pmr, Schmidhuber:2015ysx, Plehn:2022ftl}. Machine learning involves algorithms that enable computers to learn from data and make predictions or decisions without being explicitly programmed, while deep learning utilizes neural networks with multiple layers to learn complex patterns and representations from data. Convolutional Neural Networks (CNNs) are particularly effective for processing spatial data, such as images, while Long Short-Term Memory (LSTM) networks specialize in capturing temporal dependencies in sequential data \cite{Hochreiter:1997yld}.

Therefore, combining Convolutional 1D (Conv1D) layers with LSTM networks offers a powerful solution for analyzing time series data in physics. Conv1D layers effectively capture local patterns in sequential data, while LSTM networks model long-term dependencies and temporal dynamics, enabling accurate analysis of time-dependent datasets across various domains.

In this study, we apply deep learning techniques to the following example model. We consider the post-inflationary dynamics of standard model Higgs particles abundantly produced by the inflaton, where the Higgs subsequently decays into gauge bosons through $SU(2) \times U(1)$ interactions, leading to the formation of a thermal sector. During the inflaton oscillation period, numerical simulations are often necessary since the production of scalar or gauge bosons after inflation involves non-perturbative dynamics (e.g. preheating)\cite{Kofman:1994rk, Kofman:1997yn, Greene:1997fu, Rajantie:2000nj, Skullerud:2003ki,Micha:2003ws, Berges:2012ev,Figueroa:2015rqa,Enqvist:2015sua}. Since lattice simulations designed to analyze the preheating system have limitations in observing late-time dynamics up to reheating, we attempt to overcome these limitations by employing deep learning techniques.

Our paper is structured as follows: In Section \ref{setup}, we examine the tachyonic production mechanism of Higgs via its trilinear interaction with the inflaton. In Appendix \ref{lattice}, we investigate how post-inflationary $SU(2)\times U(1)$ gauge interactions are implemented on the lattice, and the simulation results will be presented in Section \ref{preheating}. We discuss the deep learning technique that we applied to our results in Section \ref{discussion} and draw conclusions in Section \ref{conclusion}. Appendix \ref{hp} provides an introduction to the hyperparameters used in deep learning.

\section{Tachyonic production of Higgs}
\label{setup}

 The minimal interactions between the inflaton $\phi$ and the Higgs $H$ are of the form \cite{Lebedev:2021xey}
\begin{equation}
\Delta V = \frac{1}{2}{m^2_{\phi}} \, \phi^2 + \frac{1}{4}{\lambda_{\phi}} \, \phi^4+ \frac{1}{2}{\lambda_{\phi h}} \, \phi^2 \, H^\dagger H+{\sigma_{\phi h}} \, \phi \, H^\dagger H - {m^2_{h}} \, H^\dagger H+{\lambda_{h}} \, (H^\dagger H)^2 \;,
\end{equation}
where we assume $\lambda_{\phi} = 0, \, \lambda_{\phi h} = 0,\,$ and $m_h=0$ for simplicity.\footnote{In the case of the quartic inflaton potential, to elevate the global minimum of the potential from negative values to zero, one should introduce an undesirable offset term \cite{Abolhasani:2009nb}.}
 In unitary gauge
\begin{equation}
H=\frac{1}{\sqrt{2}}\begin{pmatrix} 0 \\ h 
\end{pmatrix}
\, ,
\end{equation}
the potential simplifies to
\begin{equation}\label{pot3}
V=\frac{1}{2}{m^2_{\phi}} \, \phi^2  +\frac{\sigma_{\phi h}}{2} \, \phi \, h^2 +\frac{\lambda_{h}}{4} \, h^4 \, .
\end{equation}
By rewriting Eq.~\eqref{pot3} in the form
\begin{equation}\label{potshape}
V=\frac{1}{2}\left({m_{\phi}} \, \phi + \frac{\sigma_{\phi h}}{2 m_\phi} h^2 \right)^2  +\frac{1}{4} \, \left( \lambda_h - \frac{\sigma^2_{\phi h}}{2 m^2_{\phi}} \right) h^4 \, ,
\end{equation}
one finds that $\lambda_h$ must be larger than $\sigma^2_{\phi h} /2 m^2_\phi $ to make the potential bounded from below \cite{Dufaux:2006ee}. The equation of motion of $h$ is
\begin{equation}
\Box h +V_{,h}=0
\end{equation}
Quantizing $h$ with the eigenmodes $h_k(t) e^{i \bold{k} \vec{x}}$ and $\bold{k}$ the comoving momentum, one may derive linear mode equations in momentum space\footnote{The modes are rescaled to factor out the scale factor, $h_k \rightarrow a^{3/2} h_k $.}
\begin{equation}
\ddot h_k+\omega^2_k h_k=0 , \label{eomh}
\end{equation}
where
\begin{equation}
\omega^2_k = \frac{k^2}{a^2}+ \sigma_{\phi h} \phi(t) +3 \lambda_h \langle h^2 \rangle- 3 \dot a^2 /4a^2 -3 \ddot a/2a \,.
\end{equation}
 Given that negative values of the inflaton background can make the entire frequency term negative, we expect a tachyonic growth of the Higgs mode during the inflaton oscillation period. While one can theoretically examine the mathematical structure of the mode equations and quantitatively determine the number of produced particles, practical implementation proves to be challenging for various reasons \cite{Abolhasani:2009nb}. Each mode in the equation experiences movement across the instability band with varying Floquet quotient due to the expansion of space. Eq.~\eqref{eomh} becomes linearized in terms of $k$ only when we disregard the rescattering of modes with different $k$. This is not the case in the presence of the strong self-interaction and gauge interaction of Higgs, even at the onset of the inflaton oscillation period. To deal with the turbulent particle interactions, we opt to solve the equations of motion directly in spacetime through a semi-classical approach (see \cite{Cosme:2022htl} for studies on gravitational wave production in a similar framework).

\section{Post-inflationary weak interactions}
\label{preheating}
Given the gauge interactions in the standard model, Higgs particles will interact with gauge bosons upon production. Previous studies have explored scenarios where considerable Higgs condensates generate gauge bosons via resonant channels \cite{Figueroa:2015rqa, Enqvist:2015sua}. We examine a system in which field constituents, particularly focusing on the Higgs directly produced by the inflaton, interact with other fields in a turbulent manner.

    \begin{figure}[t!] 
\centering{
\includegraphics[scale=0.31]{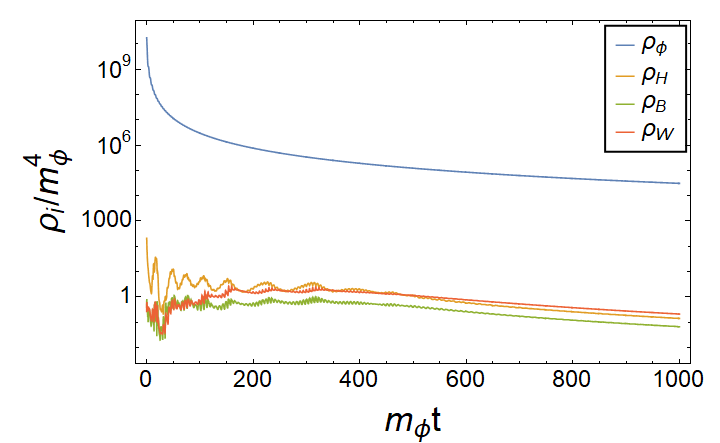}
\includegraphics[scale=0.31]{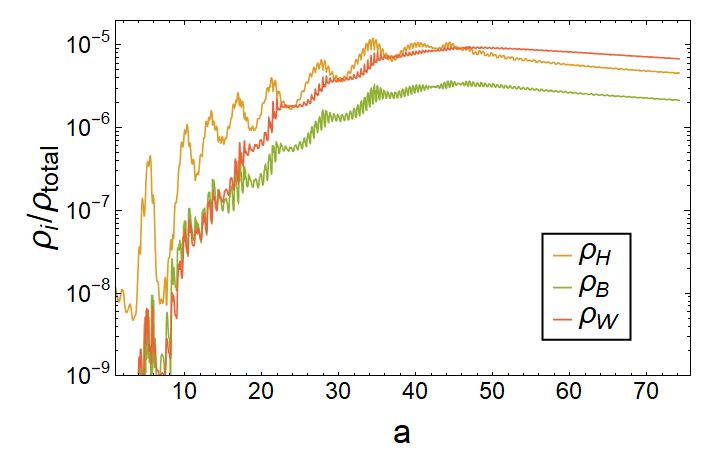}
}
\caption{ \label{energies}
Energy densities of interacting fields during preheating. {\it Left:} energy density of fields normalized by $m_\phi^4$. The $x$-axis represents time rescaled by the inflaton mass, $m_\phi$.
{\it Right:} ratio of the energy density of each field to the total energy over the scale factor.}
\end{figure}

Using a publicly available code, CosmoLattice \cite{Figueroa:2020rrl, Figueroa:2021yhd}, we conduct lattice simulations to model post-inflationary weak interactions of the Higgs (see \ref{lattice}). We set $g_2 = g_1 = 0.6$ and $\lambda_h =10^{-2}$ for the coupling constants at the inflationary physics scale to be not too different from those of the standard model. As in \cite{Cosme:2022htl}, the inflaton field starts to roll down to the minimum of potential with the bare mass $m_\phi = 1.6 \times 10^{13}$ GeV, $\phi_{init}=1 \, m_p$ and $\dot \phi_{init}= -0.71 \, m_{\phi} m_p$, where the Planck mass $m_p \approx 2.435 \times 10^{18} \text{ GeV}$. Here, $init$ refers to the initial values input into the lattice simulations. While we do not specify an inflationary model, it is assumed to converge to a quadratic inflaton potential some time after the end of inflation.\footnote{Around the end of inflation where $\phi >0$, the potential of the form Eq.~\eqref{potshape} reaches its minimum for $h=0$.} Parameters may vary within certain limits to accommodate running coupling constants or experimental constraints.

The energy densities of fields are defined
\begin{eqnarray}
\rho_{\phi} &\equiv& K_{\phi}+G_{\phi}+V(\phi) \, \\
\rho_{H} &\equiv& K_{H}+G_{H}+V(H) \, \\
\rho_{B} &\equiv& K_{U(1)}+G_{U(1)} \, \\
\rho_{W} &\equiv& K_{SU(2)}+G_{SU(2)} \, ,
\end{eqnarray}
where $V(\phi)$ and $V(H)$ represent the potential of the inflaton and Higgs doublet, respectively. Fig.~\ref{energies} shows the energy densities of fields over time for the case where $\sigma_{\phi h}=10^9 \, \text{GeV}$. The total energy density of the system is still dominated by the energy of the inflaton field, while the Higgs contributes only a small fraction of the total energy. Due to the strong self-interaction of the Higgs ($\lambda_h =10^{-2}$), tachyonic production is highly suppressed but its interaction with the inflaton is sufficiently strong to withstand the dilution by the quadratic inflaton potential ($\rho_R \propto a^{-4}$ for radiation while $\rho_{\phi} \propto a^{-3}$ for the massive inflaton field).

\begin{figure}[t!] 
\centering{
\includegraphics[scale=0.315]{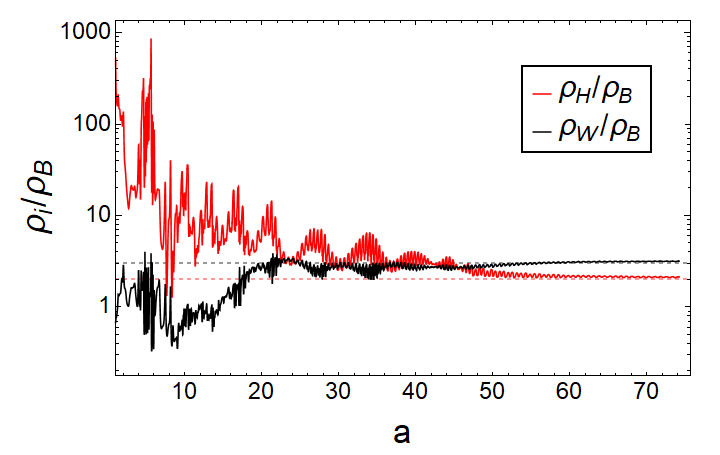}
\includegraphics[scale=0.295]{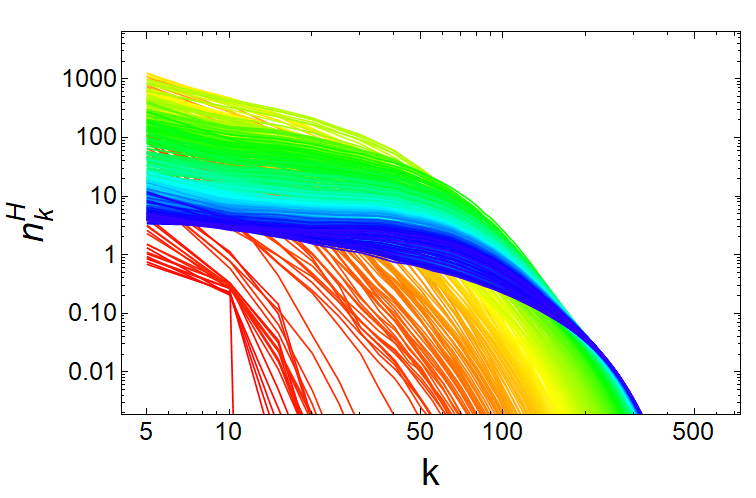}
}
\caption{ \label{prethermal}
Indication of a quasi-equilibrium state in the gauge sector. {\it Left:} energy ratio of Higgs and $SU(2)$ gauge field to $U(1)$ gauge field. The dashed grey horizontal line represents 3 which corresponds to the ratio of degrees of freedom for the $SU(2)$ gauge field to the one for the $U(1)$ gauge field ($\rho_W/\rho_B=3$). Likewise, the red dashed horizontal line represents the same quantity but for the Higgs doublet ($\rho_H/\rho_B=2$).
{\it Right:} occupation number for the Higgs doublet (red to blue over time). The comoving momentum $k$ is normalized by the inflaton mass. The mean particle distribution shifts from IR to UV as time progresses.}
\end{figure}

For a qualitative analysis, we illustrate the energy ratio of the Higgs and the $SU(2)$ gauge field to the $U(1)$ gauge field on the left of Fig.~\ref{prethermal}. We anticipate observing the equipartition of energy in the gauge sector as an indication of thermalization, in a ratio of 4:6:2 (or equivalently 2:3:1) corresponding to the degrees of freedom of the Higgs, $SU(2)$ gauge fields, and $U(1)$ gauge boson, respectively.
We have captured the moment of energy equipartition shortly after the end of inflation ($a \sim 60$ or $m_{\phi}t \sim 700$).

We excited initial particle modes for the comoving momentum in the range $5 \leq k/m_\phi \leq10$ and evolved the system in the extended range $5 \leq k/m_\phi \lesssim690$. It should be noted that our focus is on the tail end of spectra in the UV, which could potentially underestimate Higgs production in the IR. Expanding the momentum window in the IR direction leads to increased energy exchange due to the increased scattering of particles, necessitating further expansion in the UV direction as well. This results in a significant increase in simulation runtime and it is one of the key motivations for us to leverage deep learning techniques. For the subsequent discussion, we define the occupation number for the Higgs doublet as\footnote{Zero initial occupation number for scalar fields in theory corresponds to $n_{k, init}=1/2$ in our definition. For the Higgs doublet, we take into account 4 Higgs {\it d.o.f} and 1/2 for the normalization factor, resulting in $n^H_{k, init}=1/2 \times 4 \times 1/2 = 1$.}
\begin{eqnarray}
n^H_k= \sum_i n^i_k \equiv \sum_i \frac{\omega^i_k}{2} \left( | {(h^i_k)'}|^2/({\omega^i_k})^2 +  |h^i_k|^2  \right) \, ,
\end{eqnarray}
where $i$ represents the Higgs $d.o.f$ and the frequency is
\begin{eqnarray}
\omega^i_k \equiv \sqrt{k^2+a^2 \, |d^2 V(H)/d{h^i}^2|} \,.
\end{eqnarray}
The prime denotes the conformal time derivative $a (d/dt)$. For gauge bosons, since the occupation number is gauge-dependent, gauge fixing must be performed at each measurement \cite{Enqvist:2015sua,Giusti:2001xf,Bali:1994jg,Schrock:2012fj}. Due to its technical difficulty, we leave this aspect for future work.

The late-time classical reheating process of self-interacting fields or non-abelian fields exhibits a turbulent and self-similar evolution of distribution functions towards equilibrium \cite{Micha:2004bv, Berges:2013fga}. The shape of the spectra, along with the self-similar dynamics, can be comprehended within the framework of wave kinetic theory.\footnote{See \cite{Berges:2008wm, Berges:2012us} for possible IR- and UV-cascades in the context of non-thermal fixed points in scalar field theories.} This suggests the presence of a consistent trend in the evolution of the occupation number, offering a foundation for exploring the occupation number using deep learning methods.

\section{CNN-LSTM time series analysis}
\label{discussion}

The CNN-LSTM model integrates two fundamental components: the Convolutional Neural Network (CNN) and the Long Short-Term Memory (LSTM) network. Each component plays a distinct role in processing the input data to extract features and capture temporal dependencies.

The CNN component is responsible for extracting spatial features from the input data. It operates by applying convolutional filters over the input data to capture spatial patterns, such as edges, textures, and shapes. This process is facilitated by multiple convolutional layers followed by pooling layers, which help reduce spatial dimensions while retaining essential information.

On the other hand, the LSTM component specializes in modeling temporal dependencies in sequential data. It consists of stacked LSTM layers, each equipped with memory cells capable of retaining information over extended time periods. These memory cells are governed by gates that control the flow of information, enabling the model to selectively update, forget, or output information based on the input sequence.

In the parallel CNN-LSTM model, the input data are processed separately by the CNN and LSTM components. Once the CNN and LSTM components have processed their respective data streams, the output features are concatenated and fed into a linear layer (or layers) for further processing. This linear layer serves as the final step in the model, where the extracted features are transformed and mapped to the desired output space.

In our CNN-LSTM time series analysis, the input data have the shape (batch\_size, seq\_length, input\_size), which corresponds to (the number of time points, the length of sequence, the number of $k$ modes) for the preheating model that we discuss. Both the input and output contain $n_k(t)$ of the Higgs, but the model output predicts values occurring after one sequence.\footnote{In a time series context, the term `time' typically denotes the sequential order of data points rather than implying that the input variables must directly represent time intervals.}
We implement the model using PyTorch \cite{DBLP:journals/corr/abs-1912-01703}, a popular deep learning library in Python. The model architecture consists of a Conv1D layer followed by an LSTM layer. The Conv1D layer captures local patterns in the sequential data, while the LSTM layer captures long-term dependencies.

The model is trained using the training data, and the training process involves optimizing a loss function (mean absolute error) using the Adam optimizer. The training loop iterates over epochs, updating the model parameters to minimize the loss. Predictions are made for future time steps beyond the training data and compared with the actual test data. The test data are completely separated from the model training and forecasting process.

The free turbulence starts approximately around $m_\phi t \approx 400$, so we divided the data into a training set for $400<m_\phi t<1000$ and a test set for $1000<m_\phi t<1500$. We make predictions up to $m_\phi t \approx 2000$.

     \begin{figure}[t!] 
\centering{
\includegraphics[scale=0.51]{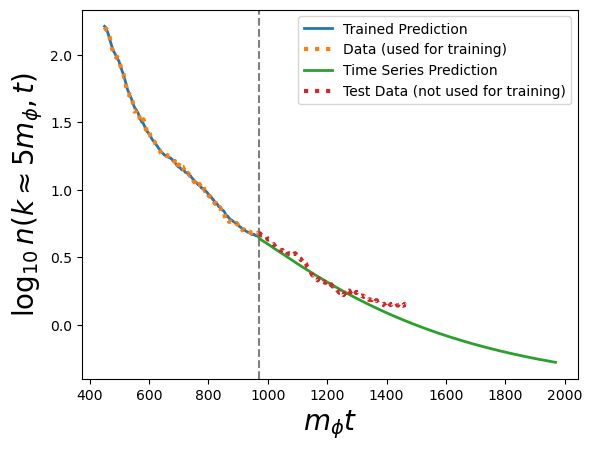}
\includegraphics[scale=0.51]{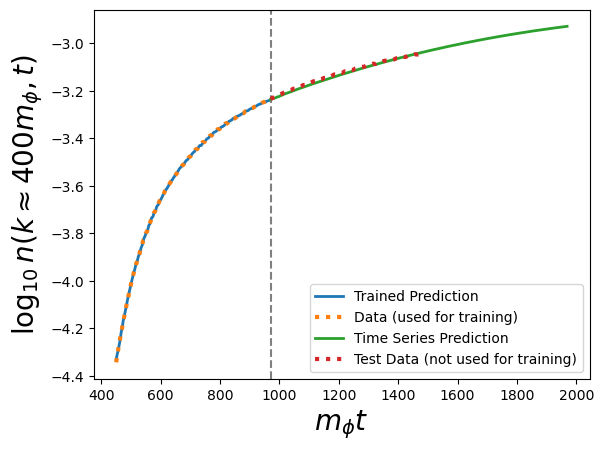}
}
\caption{ \label{pcls}
Occupation number for the Higgs field as a function of $t$ with a logarithmic scale for two $k$ modes. The gray dashed vertical line distinguishes between training data (orange dotted curve) and time series predictions (green solid curve). The trained predictions (blue solid curve) are obtained by feeding the input data into the model, while the time series predictions are recursively generated by applying the model to the output. Test data (red dotted curve) are used to calculate the validation loss. {\it Left:} $n_k(t)$ for $k \approx 5 m_\phi$. Since it corresponds to the region interacting with the inflaton, the pattern appears to be more complex.
{\it Right:} $n_k(t)$ for $k \approx 400 m_\phi$.}
\end{figure}

The model's predictions are visualized alongside the actual data as in Fig.~\ref{pcls}. It represents the actual data used for training, the trained predictions, the test data not used for training, and the predictions for future time steps for two momentum modes among 15 modes ($5 \leq k/m_\phi \lesssim400$). It is remarkable that our deep learning model managed to learn patterns and trends in the limited amount of data and make predictions similar to actual simulation results.

\begin{figure}[t!] 
\centering{
\includegraphics[scale=0.27]{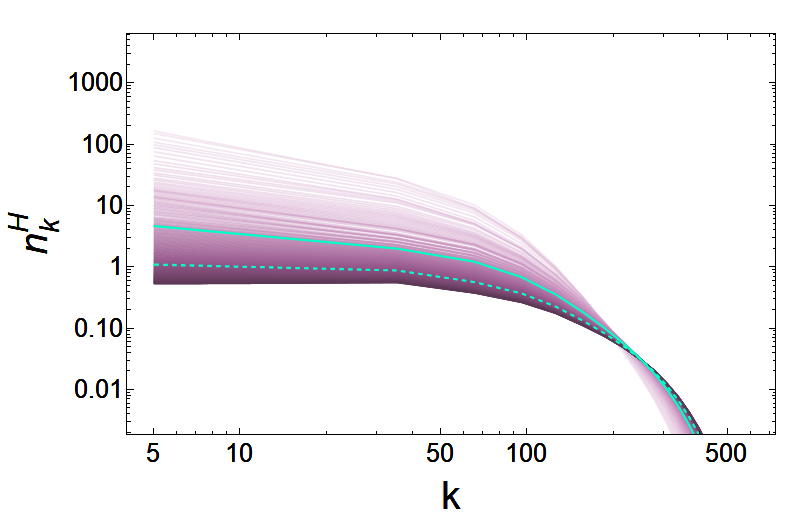}
\includegraphics[scale=0.32]{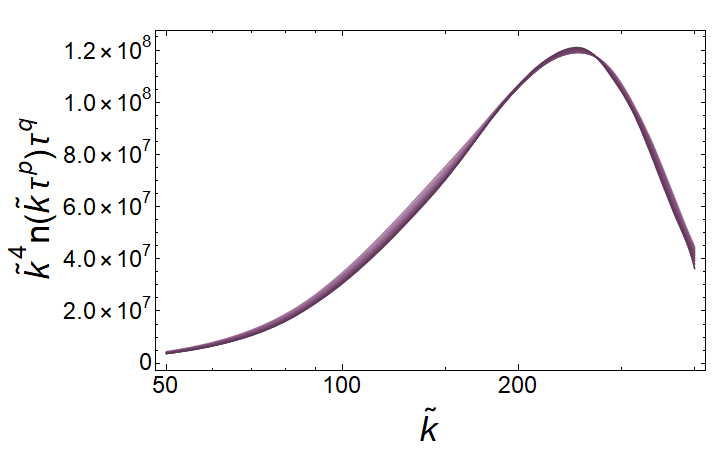}
}
\caption{ \label{pcls2}
Predicted spectra of the Higgs occupation number. {\it Left:} Occupation numbers computed for different momentum modes (15 points in total) are collected and reshaped. The darker magenta color indicates a later spectrum in time ($500 \lesssim m_\phi t \lesssim 2000$) and the cyan solid (dashed) curve represents a spectrum at $m_\phi t \approx 1000$ ($m_\phi t \approx 1500$).
{\it Right:} Self-similar form of the spectra corresponding to the left-hand side of Eq.~\eqref{eqfit}. Only predicted values outside the data range were used to fit $p$ and $q$ ($1500 \lesssim m_\phi t \lesssim 2000$).
}
\end{figure}

The predictions for each mode are collected and reshaped in Fig.~\ref{pcls2}. At the end of our predictions, around $m_\phi t \approx 2000$, it appears that we have already reached a point where the system departs from the classical regime and quantum effects become significant, indicating proximity to thermalization. By further developing the deep learning model, we can implement a more realistic cosmological model that correctly incorporates the production of the Higgs, which has been underestimated in the IR regime.

Still, we may use the predicted values to verify the self-similar evolution. Rewriting Eq.~(4) from \cite{Micha:2004bv} yields
\begin{eqnarray}
\label{eqfit}
\tilde k^4 n(k=\tilde k \tau^p, t) \tau^q = \tilde k^4 n(k=\tilde k ,t=t_0) ~,
\end{eqnarray}
where $\tilde k \equiv k \tau^{-p}$, $\tau \equiv t/t_0$, and $t_0$  denotes a fixed point in time. $p$ and $q$ represent the fitting exponents, and $p$ determines how quickly the system approaches equilibrium. Since the right-hand side of Eq.~\eqref{eqfit} is stationary, it was used as data for fitting, resulting in a good fit when $p \approx 0.16$ and $q \approx 0.63$, as illustrated in Fig.~\ref{pcls2}. For physical insights on the exponents, see \cite{Micha:2004bv, Berges:2013fga}. However, the aim of this work is not to precisely determine the exponents, but to apply deep learning techniques based on the consistent patterns in the distributions, referred to as self-similar evolution or non-thermal fixed points. The exponents may vary depending on the details of the self- or gauge interaction of the Higgs field as well as the fitting conditions. In fact, the Higgs field interacts not only through self- and gauge interactions but also with fermions, and intense fermion reactions are expected during the preheating period \cite{Garcia-Bellido:2008ycs,Berges:2013oba}. Once the code is updated to include fermions, it will be interesting to directly observe how this affects the self-evolution or non-thermal fixed points in the Higgs distribution.

\section{Conclusion}
\label{conclusion}
We consider a simple post-inflationary scenario where after inflation, the Standard Model Higgs decays into gauge bosons via weak interactions while forming a quasi-thermal state in the gauge sector. The non-perturbative nature of the system is mainly due to the $SU(2)$ gauge interaction ($g =0.6$) and Higgs self-coupling ($\lambda_h=10^{-2}$) along with the trilinear interaction ($\sigma_{\phi h} \phi H^\dagger H $ where $\sigma_{\phi h} = 10^{9} \,$GeV) between Higgs and the inflaton. Analyzing the production and decay of Higgs in this model appears challenging without relying on numerical and semi-classical approaches. Demonstrating thermalization through simulation is also very challenging, primarily because simulations are costly in practical terms and quantum effects become significant in later stages. We introduced deep learning techniques as an effort to overcome the limitations of classical lattice simulations and ultimately reproduce the thermal universe. The later stage of preheating is referred to as the regime of free turbulence, where the flow of particle numbers can be analyzed through kinetic theory, making it a suitable material for applying deep learning techniques. LSTM networks are commonly used in time series analysis, but they have limitations in predicting beyond the trained range. And since the momentum modes of the fields in the free turbulence regime have localized influence rather than behaving independently, local patterns in the momentum space can play an important role. CNNs are useful for capturing localized patterns observed in images and similar data. Therefore, we utilized a deep learning model that combines LSTM with CNN, allowing a fully connected linear layer to learn the outputs of both networks simultaneously. We applied this model to the realistic post-inflationary preheating scenario and predicted the spectra of the Higgs field beyond simulation results. While still in its early stages, this study has demonstrated its potential for advancement.

\vspace{1cm}

{\bf Acknowledgements.} J.Y. would like to thank Oleg Lebedev for useful discussions.
   This work was performed using HPC resources from the ``M\'esocentre'' computing center of CentraleSup\'elec, \'Ecole Normale Sup\'erieure Paris-Saclay and Universit\'e Paris-Saclay supported by CNRS and R\'egion \^Ile-de-France (https://mesocentre.universite-paris-scalay.fr/).     
 This project has received support from the European Union's Horizon 2020 research and innovation programme under the Marie Sklodowska-Curie grant agreement No 860881-HIDDeN, and the CNRS-IRP UCMN.

\appendix
\section{$SU(2) \times U(1)$ on the lattice}
\label{lattice}

Let us start with the $SU(2) \times U(1)$ gauge Lagrangian
\begin{equation}
\mathcal{L}_{\textrm gauge} = - \frac{1}{4}{(W^a_{\mu \nu})^2} \, - \frac{1}{4}{B^2_{\mu \nu}} \, + (D_\mu H)^\dagger(D_\mu H) + {m^2_{h}} \, H^\dagger H -{\lambda_{h}} \, (H^\dagger H)^2 \;,
\end{equation}
where $B_\mu$ is the $U(1)$ gauge boson, with $B_{\mu \nu} = \partial_\mu B_\nu - \partial_\nu B_\mu$, and $W^a_\mu$ are the $SU(2)$ gauge bosons, with $W^a_{\mu \nu}$ their field strengths. The covariant derivative of Higgs doublet $H$ is 
\begin{equation}
D_\mu H= \partial_\mu H - i g_2 W^a_\mu \tau^a H - \frac{1}{2} i g_1 B_\mu H ,
\end{equation}
where $g_2$ and $g_1$ are the $SU(2)$ and $U(1)$ couplings, respectively.

To present the equations of motion in an expanding universe, we extract useful expressions from \cite{Figueroa:2020rrl}. First, we assume a flat Friedmann-Lemaître-Robertson-Walker (FLRW) metric, characterized by the line element:
\begin{equation}
\text{d}s^2 = g_{\mu\nu}\text{d}x^\mu\text{d}x^\nu = -a(\eta)^{2\alpha}\text{d}\eta^2 + a(\eta)^2\delta_{ij}\text{d}x^i\text{d}x^j \ , \label{eq:FLRWmetric}
\end{equation}

Here, $a(\eta)$ represents the scale factor, $\delta_{ij}$ denotes the Euclidean metric, and $\alpha$ is a constant parameter chosen conveniently depending on the scenario.

The EOMs in an expanding universe are given by
\begin{eqnarray}
\phi'' - a^{-2(1 - \alpha)} {\vv\nabla}^{\,2} \hspace{-1mm}\phi + (3 - \alpha)\frac{{a'}}{a} {\phi'} &=& - a^{2 \alpha} V_{,\phi} \ , \label{eq:singlet-eom} \\
H'' - a^{-2(1 - \alpha)} {\vv D}^{\,2}H + (3 - \alpha)\frac{{a'}}{a}  {H'} &=& - \frac{a^{2 \alpha} V_{,|H|}}{2} \frac{H}{|H |} \ , 
\\
\partial_0 F_{0i} - a^{-2(1 - \alpha )}\partial_j F_{ji} + (1 - \alpha) \frac{{a'}}{a} F_{0i} &=&
a^{2 \alpha}J^B_i \ , 
\\
(\mathcal{D}_0 )_{a b} (G_{0i})^b - a^{-2(1 - \alpha )} ( \mathcal{D}_j )_{a b} (G_{ji} )^b + (1 - \alpha) \frac{{a'}}{a} (G_{0i} )^b &=& a^{2 \alpha}(J_i)_a \ , 
\\
\partial_i F_{0i} &=& a^2J^B_0 \ , \label{u1gauss}\\
(\mathcal{D}_i )_{a b} (G_{0i})^b &=& a^2(J_0)_a \,, \label{su2gauss}
\end{eqnarray}
where the prime $'$ denotes $\alpha$-time derivative $d/d\eta$ and $(\mathcal{D}_{\nu}O)_a = (\mathcal{D}_{\nu})_{a b}O_b \equiv ( \delta_{a b}  \partial_{\nu} - f_{abc} W_{\nu}^c ) O_b$.
The field strength tensors are defined as
\begin{eqnarray}
F_{\mu \nu} &\equiv & \partial_{\mu}  B_{\nu} - \partial_{\nu} B_{\mu} \ , \\
G_{\mu \nu} & \equiv & \partial_{\mu} W_{\nu} - \partial_{\nu} W_{\mu} - i g_2 [W_\mu,W_\nu] \, , 
\end{eqnarray}
and the currents are 
\begin{eqnarray}
\label{u1current}
\hspace{1.8cm} (J^B)^\mu & \equiv &  g_1\mathcal{I}m [ H^\dag (D^{\mu} H  )]\,,\\
\label{su2current}
\hspace{1.8cm} J_a^\mu & \equiv & 2g_2\mathcal{I}m [ H^{\dag} T_a( D^{\mu} H )]\,.
\end{eqnarray}
Eqs.~(\ref{u1gauss}) and (\ref{su2gauss}) are the $U(1)$ and $SU(2)$ Gauss constraints in an expanding background that can be used to check the consistency of numerical simulations.
The evolution of the scale factor obeys the Friedmann equations
\begin{eqnarray}\label{eq:FriedmannHubble}
\left({a'\over a}\right)^2 &=&  \frac{a^{2 \alpha}}{3 m_p^2}\left\langle {K}_{\phi} + {K}_{H} + {G}_{\phi}  + {G}_{H} + {K}_{U(1)} + {G}_{U(1)} + {K}_{SU(2)} + {G}_{SU(2)} + {V}\right\rangle \,,
\\
{a''\over a} &=& \frac{a^{2 \alpha}}{3 m_p^2}\left\langle (\alpha-2)({K}_{\phi}  + {K}_{H}) + \alpha({G}_{\phi}  + {G}_{H}) + (\alpha + 1)V \right.\\
&& \hspace{2cm}\left. +~ (\alpha-1)({K}_{U(1)} + {G}_{U(1)} + {K}_{SU(2)} + {G}_{SU(2)}) \right\rangle \,,\nonumber
\end{eqnarray}
where $\langle ... \rangle$ represents values averaged over space. One of the two Friedmann equations is redundant and used to check consistency as Gauss constraints. The kinetic ($K$) and gradient ($G$) energy densities are
\begin{eqnarray}
\begin{array}{lcl} \label{eq:energy-contrib}
{K}_{\phi} &=& \frac{1}{2 a^{2\alpha} } {\phi'}^2 \vspace{0.1cm}
\vspace{0.1cm}\\
{K}_{H} &=& \frac{1}{a^{2\alpha} } (D_0 H )^\dag(D_0 H)
\vspace{0.1cm}\\
{K}_{U(1)} &=& \frac{1}{2 a^{2 + 2 \alpha}}  \sum_{i} F_{0i}^2
\vspace{0.1cm}\\
{K}_{SU(2)} &=& \frac{1}{2 a^{2 + 2 \alpha}}  \sum_{a,i} (G_{0i}^a)^2
\vspace{0.1cm}\\
\end{array}
\hspace{0.1cm};\hspace{0.4cm}
\begin{array}{lcl}
{G}_{\phi} &=& \frac{1}{2 a^2} \sum_i (\partial_i \phi)^2
\vspace{0.1cm}\\
{G}_{H} &=& \frac{1}{a^2} \sum_i (D_i H)^\dag(D_i H)
\vspace{0.1cm}\\
{G}_{U(1)} &=& \frac{1}{2 a^4}  \sum_{i,j<i} F_{ij}^2
\vspace{0.1cm}\\
{G}_{SU(2)} &=& \frac{1}{2 a^4}  \sum_{a,i,j<i}  (G_{ij}^a)^2  \, . \vspace*{0.2cm}\\
\end{array}
\end{eqnarray}

One can solve the above equations in the temporal gauge with different equation-solving algorithms. We implement our model on the lattice using a publicly available code called CosmoLattice \cite{Figueroa:2020rrl, Figueroa:2021yhd}.

Program variables used for our physics model are rescaled as follows:
\begin{eqnarray}
\alpha&=&0 ~~,~~ f_* = 1.6 \cdot 10^{13} \, \text{GeV} ~~,~~\omega_* = 1.6 \cdot 10^{13} \, \text{GeV} ~~, ~~d \tau =a^{-\alpha} \omega_* dt ~~,\\ [10pt]
\tilde f &\equiv& f/f_* ~~,~~ \tilde V \equiv V/(f^2_* \omega^2_*) ~~,
\end{eqnarray}
where variables with tildes represent dimensionless program variables.
Important input parameters used for our model are set as follows
\begin{eqnarray}
\text{evolver=VV2 ~~,~~ N=160 ~~,~~ dt=0.005 ~~,~~ kIR = 5 ~~,~~ kCutOff = 10.0 \, .}
\end{eqnarray}
We evolve the field equations using the second kind of Velocity-Verlet algorithm on 160 lattice points per dimension. dt represents the time step in program unit.

\section{Hyperparameters}
\label{hp}

\begin{figure}[t!] 
\centering{
\includegraphics[scale=0.6]{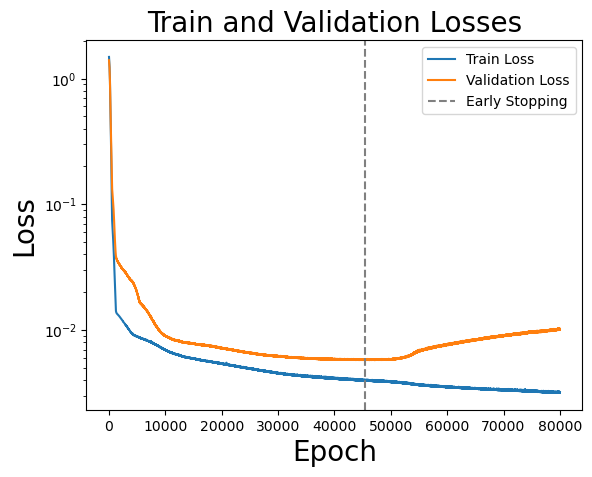}
}
\caption{ \label{losses}
Train and validation losses versus epoch. Early stopping was performed at the epoch marked by the gray dashed vertical line.
}
\end{figure}

Hyperparameters are parameters whose values are set before the learning process begins. They control aspects of the learning process itself and are not learned from the data. These parameters are external to the model and must be set based on heuristics, prior experience, or trial and error. Common hyperparameters include the learning rate, batch size, number of hidden layers, number of neurons in each layer, dropout rate, regularization strength, and optimizer choice. Proper selection and tuning of hyperparameters are crucial for achieving good performance and generalization of machine learning models.
If hyperparameters are not tuned, the results obtained through the model can either underfit or overfit the data. Therefore, it is important to compute and monitor the train loss and validation loss. Typically, the train loss decreases over epochs, so by solely examining the validation loss, we can discern whether the model is underfitting or overfitting. If the validation loss continues to decrease, it indicates underfitting, suggesting room for improvement through further training. However, if the validation loss does not decrease, it implies overfitting, where the model merely memorizes the data without understanding beyond it. Thus, it is advisable to perform early stopping at this intermediate stage, as illustrated in Fig~\ref{losses}.

\end{document}